\documentclass[12pt]{article}
\usepackage{amsmath,amssymb,geometry}
\usepackage{authblk} 
\usepackage{graphicx}
 \usepackage{booktabs}
 \usepackage{hyperref}
\usepackage[round,authoryear]{natbib} 
\title{When Small Acts Scale: Ethical Thresholds in \\ Network Diffusion}

\author{Masoud Makrehchi\thanks{Corresponding author: \texttt{<masoud.makrehchi@ontariotechu.ca>}}}
\affil{Department of Electrical, Computer \& Software Engineering,\\
Ontario Tech University, Oshawa, Ontario, Canada}
\date{}
\begin{document}
\maketitle

\begin{abstract}
Much ethical evaluation treats actions dyadically: one agent acts on one recipient.
In networked, platform-mediated environments, this lens misses how public acts diffuse.
We introduce a minimal message-passing model in which an initiating act with baseline valence $w$ spreads across a social graph with exposure $b$, per-hop salience $\alpha$, compliance $q$, and depth (horizon) $d$.
The model yields a closed-form \emph{network multiplier} relative to the dyadic baseline and identifies a threshold at $r=b\alpha q=1$ separating subcritical (saturating), critical (linear), and supercritical (geometric) regimes.
We show how common platform design levers,  reach and fan-out (affecting $b$), ranking and context (affecting $\alpha$), share mechanics and friction (affecting $q$), and time-bounds (affecting $d$), systematically change expected downstream responsibility.
Applications include pandemic mitigation and vaccination externalities, as well as platform amplification of prosocial and harmful norms.
\end{abstract}

\section{Introduction}

Much of traditional ethical analysis proceeds under two simplifying assumptions: (1) Unlimited resources: Agents are treated as if they could costlessly repeat ideal actions (time, attention, and budget constraints are abstracted away). (2) Dyadic focus: Responsibility is evaluated within an immediate agent--patient pair, largely ignoring how actions (and norms) diffuse through the surrounding social fabric.
In what follows, we bracket (1) and concentrate on relaxing (2). Our aim is to quantify how an initiating act can generate downstream effects via social propagation. 


In a dyadic setting, one agent acts on one patient. Here, “patient” means \textbf{moral patient}, the \textit{recipient} of the action, i.e., the being who can be benefited or harmed and towards whom duties are owed. It is a term from moral philosophy (not 'medical patient'). In contrast, a \textbf{ moral agent} is the actor who can be responsible for actions. 

This lens is powerful, but incomplete for modern, connected life. Social practices spread: witnesses imitate, norms update, and platforms amplify. As a result, one local act can induce a cascade of similar acts across a network. If moral evaluation ignores these spillovers, it may seriously understate or overstate --what an agent sets in motion.

Throwing a rock through a store window is a straightforward dyadic act: An agent harms a single identifiable recipient (the shop owner). In contrast, tossing a snowball near the top of a mountain can trigger an avalanche that descends, damaging homes, roads, and lives far from the point of origin. The first case is local and bounded; the second is cascading, where a small nudge unlocks a chain of connected effects. 

We develop a minimal, quantitative model for \emph{network-aware} moral responsibility. The core idea is simple: treat moral influence as a message passing on a graph, attenuating with social distance and (optionally) with imperfect compliance. This keeps the analysis transparent while making the dyadic/network contrast explicit.

After this introduction, Section~2 develops the closed-form diffusion model and characterizes its behavior at the critical threshold $r=b\alpha q=1$. Section~3 situates the framework within social psychology and normative ethics. Section~4 applies the model to a stylized worked example and two public-health cases (pandemic mitigation and vaccination), and sketches the analogy to epidemiological SIR dynamics. Section~5 details assumptions, limitations, and possible extensions. Section~6 concludes with implications for agents and for platform governance.


\section{Model}
Consider a social network viewed locally as a branching process up to depth $d\in\mathbb{N}$. An initiating agent $A$ performs a moral act of baseline weight $w>0$ toward its immediate neighbors (depth $1$). Three parameters capture network effects:
\begin{itemize}
    \item $b\ge 1$: branching factor  which is estimated by the average number of immediate neighbors per agent. 
    \item $\alpha\in(0,1]$: attenuation per hop measuring the norm strength, attention, or salience decays with distance.
    \item $q\in[0,1]$: compliance/propagation probability  which is the expected fraction who will \emph{adopt} and pass the behavior onward.
\end{itemize}

At depth $k=1,2,\dots,d$, 
we can calculate the expected number of impacted agents as follows: $C_k=b^k q^{k-1}$.\footnote{At depth $1$ the initiator directly affects $b$ agents ($q$ does not apply yet). Thereafter, each new layer forms with expected factor $qb$.}
Impact per agent which is  relative to the baseline $w$ is  $\alpha^{k-1}$.
The total (expected) network responsibility attributable to the initiator is therefore
\begin{equation}
    T(w;b,\alpha,q,d)\;=\;\sum_{k=1}^{d} \big[w\,\alpha^{k-1}\,C_k\big]
    \;=\; w\sum_{k=1}^{d} b^{k}\,(\alpha q)^{k-1}
    \;=\; w\,b\sum_{j=0}^{d-1} (b\alpha q)^j.
\end{equation}
This is a geometric series with ratio $r=b\alpha q$:
\begin{equation}
    T(w;b,\alpha,q,d)
    \;=\;
    \begin{cases}
      \displaystyle
      w\,b\,\frac{1-(b\alpha q)^d}{1-b\alpha q}, & \text{if } b\alpha q\neq 1,\\[1.2em]
      \displaystyle
      w\,b\,d, & \text{if } b\alpha q=1.
    \end{cases}
\end{equation}

\paragraph{Dyadic baseline and a network multiplier.}
If we only count the immediate layer (depth $1$), the dyadic baseline is $T_{\text{dyad}}=w\,b$. Let's define the \emph{network multiplier}
\begin{equation}
    M(b,\alpha,q,d)\;=\;\frac{T(w;b,\alpha,q,d)}{T_{\text{dyad}}}
    \;=\;
    \begin{cases}
      \displaystyle \frac{1-(b\alpha q)^d}{1-b\alpha q}, & b\alpha q\neq 1,\\[0.8em]
      d, & b\alpha q=1.
    \end{cases}
\end{equation}
This cleanly separates the local act ($w\,b$) from its network amplification ($M$). 


\subsection{Behavior and thresholds}
Let $r=b\alpha q$ denote the effective diffusion ratio and
$M_d=\sum_{j=0}^{d-1} r^j$ the network multiplier relative to the
dyadic baseline. The model yields three regimes with distinct normative
and design implications:
\paragraph{Subcritical regime ($r=b\alpha q<1$).}
When the effective ratio $r$ is below one, the network multiplier
$M_d=\sum_{j=0}^{d-1} r^j=\frac{1-r^{\,d}}{1-r}$ is bounded and
converges to $M_\infty=\frac{1}{1-r}$ as $d\to\infty$. Each additional
layer contributes $r^{k-1}$, so downstream effects decay geometrically
and most responsibility lies in the first few hops. The share captured
by the first $K$ hops is $1-r^{\,K}$; for instance, with $r=0.6$,
$K\approx 6$ captures $95\%$ of the total, while with $r=0.8$,
$K\approx 14$ suffices. Normatively, even subcritical diffusion amplifies
responsibility above the dyadic baseline by a factor of $M_\infty$, but
there is no runaway cascade. Practically, subcriticality can be secured
by reducing exposure ($b$), attenuation losses across hops ($\alpha$), or
adoption/retransmission probability ($q$).

\paragraph{Critical regime ($r=b\alpha q=1$).}
At the critical point, the network multiplier is exactly linear in depth:
$M_d=\sum_{j=0}^{d-1}1^j=d$. Responsibility neither saturates (as with
$r<1$) nor explodes (as with $r>1$); each additional hop adds the same
increment as the first hop. Consequently, the first $K$ hops capture a
$K/d$ fraction of the total. The critical line is a knife-edge: for
$r=1+\varepsilon$ with small $|\varepsilon|$, we have
$M_d=\frac{1-r^{\,d}}{1-r}=\frac{(1+\varepsilon)^d-1}{\varepsilon}
\approx d + \tfrac{d(d-1)}{2}\varepsilon$, so even tiny changes in $b$,
$\alpha$, or $q$ induce $O(d^2)$ shifts in total impact for fixed $d$.
Normatively, at criticality the dyadic baseline underestimates expected
responsibility by a factor of $d$, making horizon length (or allowable
network depth) a central design and policy lever.

\paragraph{Supercritical regime ($r=b\alpha q>1$).}
When $r>1$, the network multiplier grows geometrically:
$M_d=\sum_{j=0}^{d-1} r^j=\frac{r^{\,d}-1}{r-1}\approx \frac{r^{\,d}}{r-1}$
for large $d$. Later layers dominate: the final layer alone contributes
$\frac{r^{\,d-1}}{M_d}=\frac{(r-1)/r}{1-r^{-d}}\to (r-1)/r$, and the last
$K$ layers account for $\frac{r^{\,d}-r^{\,d-K}}{r^{\,d}-1}\to 1-r^{-K}$.
Thus supercritical diffusion concentrates responsibility near the
frontier of the cascade. Normatively, ignoring diffusion in this regime
severely understates expected responsibility. Practically, designers can
restore subcriticality by lowering exposure $b$, increasing attenuation
(losses) $\alpha^{-1}$ (i.e., reducing $\alpha$), or reducing compliance
$q$; alternatively, capping the effective depth $d$ bounds total impact
even when $r>1$.


\begin{table}[h]
\centering
\small
\begin{tabular}{@{}lllll@{}}
\toprule
Regime & Multiplier $M_d$ & As $d\to\infty$ & Dominant layers & Policy levers \\ \midrule
$r<1$ & $\frac{1-r^{\,d}}{1-r}$ & $\frac{1}{1-r}$ (bounded) & Early hops & Reduce $b$, $\alpha$, $q$ \\
$r=1$ & $d$ & --- (linear in $d$) & Even across depth & As above; cap $d$ \\
$r>1$ & $\frac{r^{\,d}-1}{r-1}$ & $\sim \frac{r^{\,d}}{r-1}$ & Late/frontier hops & Push $r<1$ or cap $d$ \\
\bottomrule
\end{tabular}
\caption{Summary of the subcritical, critical, and supercritical regimes.}
\label{tab:regimes}
\end{table}

\paragraph{Ethical upshot.}
Across all regimes, dyadic evaluation understates responsibility once
diffusion is recognized. The size of the understatement depends on $r$
and $d$: bounded for $r<1$, proportional to depth at $r=1$, and
potentially vast for $r>1$. This highlights concrete levers for design
and policy: lower exposure ($b$), increase attenuation (reduce
$\alpha$), reduce compliance ($q$), or limit effective depth ($d$).

\section{Connections to Social Psychology, Ethics, and Philosophy}
This section situates the model within two mature literatures. On the social-psychology side, decades of experiments and field studies identify regularities of influence that correspond closely to our parameters $(b,\alpha,q,d)$. On the ethics and philosophy side, debates on imperceptible harms, complicity, group agency, and institutional responsibility provide the normative scaffolding for attributing downstream effects to initiators. Together, these traditions support both the \emph{empirical plausibility} and the \emph{normative significance} of network-aware responsibility.
\begin{itemize}
\item{Mapping social influence to \texorpdfstring{$(b,\alpha,q,d)$}{(b, alpha, q, d)}:
We interpret \emph{exposure} $b$ as the number of relevant observers or contacts; \emph{salience/attenuation} $\alpha$ as the per-hop carryover of influence with social or temporal distance; \emph{compliance/propagation} $q$ as the probability that an observer adopts and passes on the behavior; and \emph{depth} $d$ as the effective number of hops over which influence remains operative. These roles provide a clean bridge from psychological mechanisms to the diffusion parameters in our model.
    \item {Evidence from conformity, social impact, and networked diffusion: }
First, the \emph{number} of observers and perceived consensus raise adoption pressure. Classic conformity experiments and modern incentivized replications show robust majority influence even under accuracy incentives \citep{asch1951,franzen2023}; in our terms, higher perceived agreement increases effective $q$, while larger audiences increase $b$. Second, Latané’s \emph{social impact theory} models influence as a multiplicative function of \emph{strength}, \emph{immediacy}, and \emph{number}; the decline with distance (immediacy) parallels our hop-wise $\alpha$ \citep{latane1981}. Third, the \emph{structure} of ties matters: behaviors requiring reinforcement travel farther and faster in clustered networks than in random ones \citep{centola2010}, effectively raising $\alpha q$ because multiple nearby confirmations sustain uptake.
\item {Spillovers and boundary conditions}
Randomized public-goods experiments find that cooperative (and uncooperative) behavior \emph{cascades} and remains detectable up to three degrees of separation \citep{fowler2010}, supporting responsibility that extends beyond the initiating dyad as depth $d$ grows. Field experiments using \emph{descriptive norms} (e.g., hotel towel reuse) reliably increase compliance \citep{goldstein2008}, providing an applied lever on $q$. At the same time, bystander studies show \emph{diffusion of responsibility}: as the number of bystanders grows, the probability that any one person intervenes can fall \citep{darley1968}, indicating contexts where $q$ may \emph{decrease} with $b$. Finally, source strength matters: obedience results indicate that high-authority actors can dramatically elevate compliance \citep{milgram1963}, functioning as high-impact initiators in our framework (raising effective $q$ and slowing attenuation).
\item{Interpretation: }
These findings give psychological content to $(b,\alpha,q,d)$ and empirically ground the threshold $r=b\alpha q$: when $r>1$, cascades are expected; when $r<1$, diffusion saturates. They also warn that the parameters are \emph{context-sensitive} and sometimes \emph{non-monotonic} (e.g., the bystander effect), which underscores why institutional design targeting $b,\alpha,q,d$ is consequential.

\item{Connections to ethics and philosophy: }
The cascade model resonates with several established debates. Work on collective action and \emph{imperceptible harms} (Parfit’s “Harmless Torturers”; Kagan; Nefsky) explains why moral reasons persist even when individual contributions seem inefficacious; our model makes this precise by showing how small acts can cross thresholds through diffusion \citep{parfit1984,kagan2011,nefsky2019}. Accounts of \emph{complicity} and \emph{collective responsibility} (Kutz; French; List $\&$ Pettit; Isaacs) justify attributing responsibility within groups and organizations; $(b,\alpha,q,d)$ then provide a tractable way to estimate expected downstream effects inside such \emph{group agents} \citep{kutz2000,french1979,listpettit2011,isaacs2011}. Institutional diffusion and the \emph{problem of many hands} (Thompson) motivate design levers that lower $b$, attenuate $\alpha$, or reduce $q$ (via friction, transparency, or counter-speech), thereby shrinking the network multiplier and clarifying accountability in complex systems \citep{thompson1980,thompson2017designing}. Finally, social-norms theory (Bicchieri) and demandingness arguments (Singer) suggest that distance and numerosity need not dilute duty; when $b\alpha q>1$, universalization predicts large expected externalities of local acts, strengthening reasons to set good precedents and refrain from harmful ones \citep{bicchieri2006,singer1972}.
}
\end{itemize}
Network-aware responsibility is not speculative. It rests on replicated regularities of social influence and dovetails with normative theories that already treat shared, thresholded, and institutionally mediated outcomes as morally salient. Practically, when $(b,\alpha,q,d)$ are high (or can be made high by design), small local acts by high-reach agents predictably scale and should be evaluated, and governed, accordingly.

\section{Examples and applications}

In this section, we present three illustrative cases. First, a stylized worked example clarifies the mechanics of the model. Second, we consider pandemic mitigation (e.g., masking and distancing) as a moral act. Third, we analyze vaccination uptake and its externalities as a moral act.

\label{regimes}
\subsection{Worked example}
Set $w=1$, $q=1$, $\alpha=\tfrac{1}{2}$, $b=5$, $d=7$. Then $b\alpha=2.5$ (supercritical). The total responsibility is
\begin{align}
    T &= b\,\frac{1-(b/2)^d}{1-b/2}
      = 5\,\frac{1-2.5^7}{1-2.5}
      = 2031.171875 \;\approx\; 2031.17,
\end{align}
while the dyadic baseline is $T_{\text{dyad}}=b=5$. The multiplier is
\begin{equation}
    M \;=\; \frac{T}{b}
      \;=\; \frac{2031.171875}{5}
      \;=\; 406.234375 \;\approx\; 406.23.
\end{equation}
So the network-aware responsibility is over $400\times$ the dyadic count in this setting.

\subsection{Pandemic mitigation (masking/distancing as moral acts)}
Consider early-pandemic norms where an agent models masking/distancing that others may adopt after social reinforcement.
Let $w=1$ (per-person benefit unit), $b=8$ (relevant daily contacts), $\alpha=0.7$ (norm salience decays with distance), $q=0.6$ (probability of adopting and passing on), $d=5$ (roughly a workweek of indirect influence).
Then $r=b\alpha q=8\times 0.7\times 0.6=3.36>1$ (supercritical):
\[
T = 8\,\frac{1-3.36^{\,5}}{1-3.36} \approx 8\,\frac{1-427.9}{-2.36}
\approx 8\times 181.1 \approx 1448.8,\quad
M\approx \frac{1448.8}{8}\approx 181.1.
\]
Counting only direct effects ($T_{\text{dyad}}$) misses two orders of magnitude of downstream benefit. Responsibility for \emph{failing} to model/encourage mitigation (take $w<0$) scales the same way in magnitude.

\subsection{Vaccination externalities (herd immunity and network duty)}
Suppose an agent publicly endorses vaccination, nudging peers to schedule a dose.
Take $w=1$, $b=5$, $\alpha=0.6$ (messages retain much of their force across close ties), $q=0.7$, $d=6$.
Then $r=2.1>1$:
\[
T = 5\,\frac{1-2.1^{\,6}}{1-2.1}
= 5\,\frac{1-85.77}{-1.1}
\approx 5\times 77.06 \approx 385.30,\quad
M\approx 77.06.
\]
If institutional friction lowers salience $\alpha$ to $0.3$ and compliance to $q=0.4$, then $r=0.6<1$:
\[
M \to \frac{1}{1-0.6}=2.5,\qquad T\approx 12.5 \text{ (vs. } T_{\text{dyad}}=5\text{)}.
\]
Even subcritical settings amplify responsibility above dyadic counts, while small improvements in $\alpha$ or $q$ (e.g., easy booking links, trusted messengers) can tip the system supercritical. For \emph{anti-vaccination} messages, set $w<0$: the same multipliers quantify expected harm.

\subsection{Notes on mapping to epidemiology}


Our effective ratio $r=b\,\alpha\,q$ is a behavioral analogue of a reproduction number: when $r>1$, cascades are expected; when $r<1$, diffusion dies out, just as in SIR models an outbreak grows only if $R_0>1$ \cite{hethcote2000sir}. Network heterogeneity (e.g., high-degree hubs) can raise the effective $r$ by increasing exposure and retention, mirroring how heavy-tailed degree distributions lower epidemic thresholds in contagion models. The parallel is conceptual, our $r$ governs norm diffusion rather than pathogen biology, but it clarifies why seemingly small acts by high-reach agents can carry outsized moral weight.

For comparison, recall the classical SIR model, which partitions a well-mixed population $N$ into $S(t)$ (susceptible), $I(t)$ (infectious), and $R(t)$ (removed), with transmission rate $\beta$ and recovery rate $\gamma$:
\[
\frac{dS}{dt} = -\beta \frac{S I}{N},\qquad
\frac{dI}{dt} = \beta \frac{S I}{N} - \gamma I,\qquad
\frac{dR}{dt} = \gamma I.
\]
The basic reproduction number is $R_{0}=\beta/\gamma$; early growth requires $R_{0}>1$ (with $S(0)\approx N$). Common variants include SIS (no lasting immunity), SEIR (latent “Exposed” stage), and network SIR models that replace well-mixing with a contact graph.

\section{Discussion}
This section draws out the normative implications of the model, clarifying whose interests are counted, how platform and policy levers map to the parameters $(b,\alpha,q,d)$, and how the framework relates to the Kantian idea of universalization.
\begin{itemize}
    \item {Who counts as impacted?}
The initiator is credited (or charged) with both the direct effect and the \emph{expected} downstream effects produced by imitation, norm uptake, or platform amplification. This supplements, rather than replaces, local duties by adding a transparent spillover term to dyadic evaluation.
\item {Policy and design:}
When exposure $b$ is large (e.g., broadcast media, social platforms) or when salience $\alpha$ and compliance $q$ are high (clear norms, trust, strong incentives), the product $r=b\alpha q$ may exceed $1$, making small local acts morally large in expectation. This strengthens duties for high-reach agents (leaders, influencers, institutions) and motivates design levers that reduce $b$ (rate limits, audience caps), reduce $\alpha$ (context labels, demotion, friction), reduce $q$ (share friction, verification, deliberation prompts), or cap $d$ (time-to-live, recency windows).
\item {Universalization as diffusion:}
Kant’s universalization test is a deontic constraint on maxims, not a sociological prediction; nonetheless, in connected settings many maxims are public and imitable. The model links the counterfactual “what if everyone acted like this?” to the empirical ratio $r=b\alpha q$ that governs spread. When $r>1$, publicly acting on a maxim is not merely a thought experiment—it is likely to initiate diffusion; when $r<1$, the test still applies normatively, but uptake is expected to saturate. This explains why precedent-setting by high-reach agents can carry outsized moral weight.

\end{itemize}

When $r>1$, the expected number of adopters grows with depth (Sec. \ref{regimes}). In such contexts, universalization is not merely hypothetical:
publicly acting on a maxim approximates initiating a trajectory toward
its broader uptake. Formally, the total expected externality of setting
a precedent with baseline valence $w$ is
\[
T = w\,b\sum_{j=0}^{d-1} r^j
= 
\begin{cases}
w\,b\,\dfrac{1-r^{\,d}}{1-r}, & r\neq 1,\\[0.6em]
w\,b\,d, & r=1.
\end{cases}
\]
If $w>0$ and $r>1$, universalization and diffusion align to amplify the
good; if $w<0$ and $r>1$, they align to amplify the harm, strengthening
pro tanto reasons to refrain. When $r<1$, diffusion saturates; the
universalization test retains its normative role, but it is less
predictive of actual uptake.

\emph{Design implication.} Institutions can shape the link between
universalization and real trajectories by adjusting $b$ (visibility and
reach), $\alpha$ (per-hop salience), and $q$ (compliance). For good
maxims (e.g., prosocial norms), raise these parameters; for bad ones
(e.g., harmful misinformation), lower them or cap the effective depth
$d$.


\subsection{Assumptions, limits, and extensions}
The formalism is intentionally simple. Here we note the main idealizations and how to relax them without changing the core claims.

\begin{itemize}
  \item \textbf{Network structure (tree vs.\ graph).} Real networks have cycles and reconvergent paths, so the branching view can overcount. Replace the tree with a weighted adjacency matrix $A$ and sum $\sum_{j=0}^{d-1} (\alpha q A)^j$; the infinite-horizon limit exists when $\rho(\alpha q A)<1$.

If we simply count how many agents are reached (ignoring attenuation), the expected total up to depth $d$ is
\[
    N(d;q) \;=\; \sum_{k=1}^{d} b^{k} q^{k-1}
    \;=\; b \sum_{j=0}^{d-1} (b q)^j
    \;=\; 
    \begin{cases}
      \displaystyle b\,\frac{1-(b q)^d}{1-b q}, & b q\neq 1,\\[0.8em]
      \displaystyle b\,d, & b q=1.
    \end{cases}
\]
Setting $q=1$ recovers $N=\sum_{i=1}^{d} b^{i}=\frac{b(b^{d}-1)}{b-1}$.

  \item \textbf{Parameter homogeneity.} We treated $b$, $\alpha$, and $q$ as constants. In practice they vary by node, tie, and context. Allow $b_i$, $\alpha_{e}$, $q_i$ (or depth-dependent $\alpha_k$, $q_k$); the analysis proceeds with the same series using effective ratios determined by network structure (e.g., via $\rho(\alpha q A)$).

  \item \textbf{Resource limits and horizons.} Time, attention, and budget constraints can be modeled by letting $\alpha_k$ or $q_k$ decay with depth (e.g., $\alpha_k=\alpha\,g(k)$ with $g(k)\!\downarrow\!0$), or by explicitly capping the horizon $d$ (platform TTL, recency windows).

  \item \textbf{Nonlinear response and saturation.} Real effects plateau. Replace the per-hop factor $\alpha^{k-1}$ with a concave response (e.g., $f(k)$ with $f'(k)\!\downarrow\!0$) or introduce per-layer saturation/capacity constraints. The qualitative regime distinction (subcritical/critical/supercritical) remains informative for the local dynamics.

  \item \textbf{Valence (benefit vs.\ harm).} The machinery applies symmetrically to benefits and harms; the baseline weight $w$ may be positive or negative, and mixed channels can be handled additively.
\end{itemize}

\section{Conclusion}
Classical dyadic evaluation understates the moral significance of acts in connected environments. This paper introduced a minimal diffusion model that makes that understatement explicit. By parameterizing exposure ($b$), per-hop salience ($\alpha$), compliance ($q$), and horizon ($d$), we derived a closed-form \emph{network multiplier} and identified a threshold at $r=b\alpha q=1$ that separates saturating, linear, and cascading regimes. The result is a tractable way to connect normative assessment to concrete features of contemporary social and technological systems.
\begin{itemize}
    \item{What the model clarifies:}
First, it shows \emph{how much} dyadic accounting misses, and \emph{why}: in subcritical settings ($r<1$), most responsibility concentrates in the first few hops and the miss is bounded; at criticality ($r=1$), responsibility scales with depth; in supercritical contexts ($r>1$), later layers dominate and the miss can be orders of magnitude. Second, it reframes precedent-setting as high leverage. When $r>1$, publicly acting on a maxim is not merely a thought experiment about universal law; it is a plausible trajectory toward wider uptake. Third, it distinguishes \emph{local duties} from \emph{spillover duties}: accounting for downstream effects supplements, rather than replaces, first-hop obligations.
\item{Design and governance implications:}
Because platform and institutional choices directly move $(b,\alpha,q,d)$, the model provides a clear menu of levers. To curb harmful cascades, reduce exposure ($b$) through fan-out limits or audience caps, dampen per-hop salience ($\alpha$) via ranking and contextualization, decrease compliance ($q$) with friction and verification, or cap horizons ($d$) by limiting time-to-live and recency windows. Conversely, to promote beneficial norms, the same levers can be tuned in the opposite direction. A general principle emerges: \emph{responsibility and design should be cascade-sensitive}. High-reach agents and high-amplification systems carry stronger pro tanto duties because small local choices predictably scale.

\item{Scope and limits:}
The formalism is intentionally spare. We idealized network structure as branching; a graph formulation using a weighted adjacency matrix $A$ corrects overcounting from cycles and reconvergence and yields the same intuition via $\rho(\alpha q A)$. We treated parameters as homogeneous; heterogeneous $b_i$, $\alpha_{e}$, or $q_i$ (or depth-dependent schedules) can be accommodated with effective ratios. We evaluated expected impact ex ante; this avoids moral luck from rare outliers but shifts attention to measurement: estimating $(b,\alpha,q,d)$ in situ is an empirical task. None of these refinements alter the central message: once diffusion is visible, dyadic evaluation is systematically incomplete.

\item{A research agenda:}
Several directions follow naturally. (i) \emph{Measurement}: develop experimental and observational strategies to estimate $r$ in specific domains (e.g., platform A/B tests for $b$ and $q$, field experiments or audit studies for $\alpha$). (ii) \emph{Heterogeneity and structure}: quantify how hubs, clustering, and assortativity shift effective $r$ and redistribute responsibility across positions in a network. (iii) \emph{Apportionment}: extend the model with principled schemes (e.g., Shapley-style attributions) to divide expected downstream effects among multiple initiators. (iv) \emph{Nonlinear response}: incorporate saturation and capacity limits at each hop and study how these alter regime boundaries. (v) \emph{Policy evaluation}: use the framework ex ante to compare governance options that target $b$, $\alpha$, $q$, or $d$ under uncertainty and normative risk.
\end{itemize}
A great deal of contemporary moral life unfolds on networks designed for rapid diffusion. The simple ratio $r=b\alpha q$ turns abstract talk of “influence” into an analyzable quantity, and the three-regime structure translates directly into guidance for agents and designers. Viewed this way, many disagreements about responsibility are not merely theoretical; they are disagreements about the values of $(b,\alpha,q,d)$ in actual contexts, and about who should bear the obligation to move them. Making those parameters explicit helps align ethical evaluation with social reality—and equips institutions to prevent harmful cascades while enabling prosocial ones.


\bibliographystyle{plainnat}
\bibliography{reference}

\end{document}